\documentclass{ws-p10x7}
\input epsf.tex
\def\DESepsf(#1 width #2){\epsfxsize=#2 \epsfbox{#1}}

\begin{document}

\title{Yukawa Textures And Horava-Witten M-Theory}

\author{R. Arnowitt and B. Dutta}

\address{Center For Theoretical Physics, Department of Physics, Texas A$\&$M
University, College Station TX 77843-4242}

\twocolumn[\maketitle\abstract{The general structure of the matter Kahler
metric in the $\kappa^{2/3}$ expansion of Horava-Witten M-theory with nonstandard
embeddings is examined. It is  shown that phenomenological models based on this
structure can lead to Yukawa and V$_{\rm CKM}$ hierarchies (consistent with all
data) without introducing ad hoc small parameters if the 5-branes lie near the
distant orbifold plane and the instanton charges of the physical plane vanish.
M-theory thus offers an alternate way of describing these hierarchies,
different from the conventional models of Yukawa textures. }]
\section{Introduction}Over the past year considerable progress has been made in
understanding Horava-Witten heterotic M-theory \cite{hw1} with ``non-standard"
embeddings For a review, see \cite{o}. In this picture, space has an 11
dimensional orbifold structure of the form (to lowest order) $M_4\times
X\times S^1/Z_2$ where $M_4$ is Minkowski space, $X$ is a 6 dimensional (6D)
Calabi-Yau space, and $-\pi\rho\leq x^{11}\leq\pi \rho$. The space thus has two
orbifold 10D manifolds $M_4\times X$ at the $Z_2$ fixed points at $x^{11}=0$
and $x^{11}=\pi\rho$ where the first is the visible sector and the second is
the hidden sector, each with an a priori $E_8$ gauge symmetry. In addition
there can be a set of 5-branes in the bulk at points
$0< x_n<\pi\rho$, $n=1...N$ each spanning $M_4$ (to preserve Lorentz
invariance) and wrapped on a holomorphic curve in $X$ (to preserve N=1
supresymmetry).

In general, physical matter lives on the $x^{11}=0$ orbifold plane and only
gravity lives in the bulk. The existence of 5-branes allows one to satisfy the
cohomological constraints with $E_8$ on the $x^{11}=0$ plane breaking to $G\times H$
where $G$ is the structure group of the Calabi-Yau manifold and $H$ is the
physical grand unification group. Thus  non standard embeddings allow naturally
for physically interesting grand unification groups. We consider here the case
$G=SU(5)$, and hence $H=SU(5)$.

Phenomenology has played an important role in guiding the general structure of
Horava-Witten theory. Thus the Calabi-Yau manifold is assumed to have a size
$r^{-1}\simeq M_G$ to account for the success of grand unification at
$O(10^{16}$ GeV) and this then requires the orbifold scale to be
$(\pi\rho)^{-1}\simeq 10^{15}$ GeV, to account for the size of the 4D Planck
mass. In addition, constructing three generation models has been an important
element in string theory from its inception, and recently, three generation
models with a Wilson line breaking $SU(5)$ to $SU(3)\times SU(2)\times U(1)$
have been constructed in the M-theory frame work using torus fibered Calabi-Yau
manifolds (with two sections)\cite{dopw}. Also, the general structure (to the
first order) of the Kahler metric of the matter field has been
constructed\cite{low4}. We examine here  this structure and show that it can lead
to Yukawa textures with all CKM and quark mass data in agreement with experiment, 
without any undue fine
tuning. Thus M-theory leads to a new way of considering the Yukawa sector of the
Standard Model. It is also possible to show that a three generation model with a
Wilson line (to break SU(5) to the standard model) and possessing some of the
basic properties needed for the phenomenology exists for the Calabi-Yau manifold
with del-Pezzo base $dP_7$ \cite{ad}.

This note is a summary of the above results, and details can be found in
\cite{ad}

\section{Kahler Metric} The bose part of the 11 dimensional gravity multiplet
consists of the metric tensor $g_{IJ}$, the antisymmetric 3-form $C_{IJK}$ and
its field strength $G_{IJKL}=24
\partial_{[I}C_{JKL]}$. ($I,\,J,\,J,\,K=1...11.$).
The $G_{IJKL}$ obey field equations $D_IG^{IJKL}=0$ and Bianchi identities
\begin{eqnarray}\nonumber(dG)_{11RSTU}&=&4\sqrt 2\pi({\kappa\over
{4\pi}})^{2/3}[J^0\delta(x^{11})\\\nonumber+J^{N+1}\delta(x^{11}-\pi \rho)
&+&{1\over
2}\Sigma^N_{n=1}J^{n}(\delta(x^{11}-x_n)\\&+&\delta(x^{11}+
x_n))]_{RSTU}
\label{eq3}\end{eqnarray}
Here $(\kappa^{2/9})$ is the 11 dimensional Planck scale, and $J^{n}$,
$n=0,\,1,\,...N+1$ are sources from orbifold planes and the $N$ 5-branes.
These equations can be solved perturbatively in powers of $(\kappa^{2/3})$
\cite{low4}. The effective 4D theory can then be characterized by a Kahler
potential $K=Z_{IJ}{\bar C^I}C^J$, Yukawa couplings $Y_{IJK}$ for the matter
fields $C^{I}$ and gauge functions from the physical orbifold plane $x^{11}$=0.
To first order, $Z_{IJ}$ takes the form \cite{low4}. 
\begin{equation}  Z_{IJ}=e^{-K_T/3}[G_{IJ}-{\epsilon\over{2V}}\tilde{\Gamma}^i_{IJ}
\Sigma^{N+1}_0(1-z_n)^2\beta^{(n)}_i]
\label{eq2}\end{equation}Here $\epsilon=(\kappa/4\pi)^{2/3}2\pi^2\rho/V^{2/3}$ is the
expansion parameter. $V$ is the Calabi-Yau volume, $G_{IJ}$, $\tilde\Gamma^i_{IJ}$ and
$Y_{IJK}$ can be expressed in terms of integrals over the Calabi-Yau manifold
\cite{low4}, and $K_T$ is the Kahler potential for the moduli.
\vspace{-0.3cm} 
\section{Phenomenological Yukawa Matrices}If the perturbation analysis is to be a
reasonable approximation, the second term of Eq.(2) should be a small
correction. A priori one expects $G_{IJ}$, $\tilde\Gamma^{i}_{IJ}$ and
$Y_{IJK}$ to be characteristically of $O(1)$, and the parameter $\epsilon$ is
not too small. However, the second term will be small if $\beta^{(0)}_i$ were
to vanish and if the 5-branes were to be near the distant orbifold plane i.e.
$d_n\equiv1-z_n$ were small, where $z_n=x_n/{\pi\rho}$. In the following we will assume then that
\begin{equation}\beta^{(0)}_i=0;\,\,d_n=1-z_n\cong0.1\end{equation}
The condition $\beta^{(0)}_i=0$ is non trivial, but it is possible to show that
a three generation model of a torus fibered Calabi-Yau manifold with Wilson like
breaking $SU(5)$ to $SU(3)\times SU(2)\times U(1)$ with del -Pezzo
base $dP_7$ has this property\cite{ad}.

Eq.(3) then suggests that it is the $\epsilon$ term of Eq.(2) that are the
third generation contributions to the Kahler metric. A simple phenomenological
example for the u and d quark contributions with these properties (and containg
the maximum numbers of zeros) is $(f_T\equiv exp(-K_T/3))$:
\begin{eqnarray}\nonumber Z^u&=&f_T\left(\matrix{ 1  & 0.345  & 0 \cr 0.345 & 0.132 & 0.639 d^2
\cr 0  & 0.639 d^2& 0.333 d^2 }\right); \\
Z^d&=& 
f_T\left(\matrix{ 1 & 
0.821  & 0 \cr 0.821 & 0.887 & 0 \cr 0  & 0& 0.276  }\right).
\label{eq27}\end{eqnarray}
with Yukawa matrices ${\rm diag}Y^u$={(0.0765, 0.536, 0.585 $Exp[
\pi i/2]$)} and ${\rm diag}Y^d$={(0.849,\, 0.11,\, 1.3)}. 

These expression offer an alternate possibility for generating Yukawa
hierarchies. Thus to obtain the physical Yukawa matrices, one must first
diagonalize the Kahler metric and then rescale it to unity. Then using the
renormalization group equations, one can generate the CKM matrix, and the quark
masses. The results are given in the following table:
\begin{center} \begin{tabular}{|c|c|c|}  
 \hline Quantity&Th. Value&Exp. Value\cite{com}\\\hline
$m_t$(pole)&170.5& 175$\pm$ 5\\
$m_c$($m_c$)&1.36&1.1-1.4\cr
$m_u$(1 GeV)& 0.0032&0.002-0.008\\
$m_b$($m_b$)& 4.13&4.1-4.5\cr
$m_s$(1 GeV)& 0.110&0.093-0.125\cite{sa}\\
$m_d$(1 GeV)& 0.0055&0.005-0.015\\
$V_{us}$&0.22&0.217-0.224\\
$V_{cb}$&0.036&0.0381$\pm$0.0021\cite{ss}\\
$V_{ub}$&0.0018&0.0018-0.0045\\
$V_{td}$&0.006&0.004-0.013\\
\hline\end{tabular}\end{center}\vspace{0.3cm}
and sin$2\beta$=0.31 and sin$\gamma$=0.97. The agreement with experiment is
quite good. Note also that $m_u/m_d$=0.582 and $m_s/m_d$=20.0 in good agreement
with Leutwyler evaluations\cite{leut} 0.553$\pm$ 0.043 and 18.9$\pm$0.8.

 While the precise choice of entries in $Z^{u,d}$ and $Y^{u,d}$ are
chosen to obtain the above results, as shown in \cite{ad}, the quark mass
hierarchies arise naturally from a Kahler metric of the type of Eq. (4).
Similarly the smallness of the off diagonal $V_{CKM}$ matrix elements also occur
 naturally as a
consequence of the above model. Thus it is possible for  M-theory to generate
the Yukawa hierarchies without any undue fine tuning and without introducing
ad hoc very small off diagonal entries.

\section{Conclusion} Horava-Witten M-theory has now progressed to the point
where it offers a fundamental framework for building phenomenological models.
Thus it allows for conventional GUT groups (e.g. $SU(5)$, $SO(10)$),
accommdates grand unification at $M_G\cong3\times 10^{16}$ GeV, and has three
generation models where the GUT group breaks to the Standard Model at $M_G$ by
a Wilson line.

M theory with non-standard embeddings also offers new possibility of encoding
the Yukawa hierarchies in the Kahler metric. This can happen naturally if the
5-branes cluster near the hidden orbifold plane ($d_n\equiv1-z_n\simeq0.1$) and
the instanton charges of the physical plane vanish ($\beta^{(0)}_i=0$). It is
possible to construct three generation manifolds possessing these
properties\cite{ad}. Then if the $\kappa^{2/3}$ term of the Kahler metric is
attributed to the third generation of the u-quarks, one can construct models
possessing all the experimental hiererchies without any undue fine tuning or ad
hoc small parameters. While models of this type are to be viewed as ``string
inspired" as one can not perform the integrals over the Calabi-Yau manifold, 
they may give 
general insights into the nature of the Calabi-Yau manifold.

\section{Acknowledgement}This work was supported in part by NSF grant no.
PHY-9722090.

\end{document}